# My Life and My Journey through Solar Physics

## Arnab Rai Choudhuri

[On 18 February 2018 – one day before the conference *International Astronomical Union Symposium 340: Long-Term Datasets for the Understanding of Solar and Stellar Magnetic Cycles* (19 – 24 February 2018) – a one-day workshop on *Solar-Stellar Magnetism: Past, Present and Future* was organized at the same venue (B.M. Birla Auditorium, Jaipur, India). According to the official website of this workshop:

> This one day workshop is being organized in India on the occasion of Arnab Rai Choudhuri's 60th year to discuss outstanding challenges and future prospects in our understanding of the origin of solar-stellar magnetism – a topic in which Choudhuri has made significant contributions. The scientific sessions will be structured around invited discourses and a panel discussion. The workshop will precede (by one day) the International Astronomical Union Symposium 340 (Long-Term Datasets for the Understanding of Solar and Stellar Magnetic Cycles) and will be held in the same venue.

The Organizing Committee consisted of Dibyendu Nandi, IISER Kolkata (SOC Chair); Dipankar Banerjee, IIA; Ramit Bhattacharyya, PRL; Piyali Chatterjee, IIA; Prateek Sharma, IISc; Prasad Subramanian, IISER Pune; Durgesh Tripathi, IUCAA.

I was asked to give a 45-minute talk on *My Life and My Journey through Solar Physics*. Since I have never given such a talk and I was somewhat nervous about it, I wrote down the talk while preparing it – something which I normally would not do for an astrophysics talk. What follows is a polished version of the write-up.]

**Solar-Stellar Magnetism: Past, Present and Future**
18 February, 2018
B.M. Birla Auditorium, Jaipur, India
http://www.cessi.in/ssm/

**PROGRAM**

**09.30 – 10.00: Registration (on-site)**

10.00 – 10.35: Magnetic Fields in the Universe (Kandaswamy Subramanian)
10.35 – 11.10: Stellar Coronae and Super Flares: Solar Implications (Kazunari Shibata)

**11.10 – 11.30: Tea/Coffee Break**

11.30 – 12.05: Solar Magnetism: Challenges for Solar Cycle Models (Robert Cameron)
12.05 – 12.40: Helioseismic Perspective of Solar Cycle Related Changes in Sun (Sarbani Basu)
12.40 – 13.15: Helicity Constraints on Solar Activity (Alexei Pevtsov)

**13.15 – 14:15: Lunch Break**

14.15 – 14:50: Dynamo Models of the Solar Cycle (Allan Sacha Brun)
14.50 – 15:25: Magnetic Flux Tube Instabilities (Kristof Petrovay)
15:25 – 16:00: Flux Tube Dynamics in the Solar Atmosphere (Vasilis Archontis)
16.00 – 16:45: Panel Discussion on Outstanding Issues (Convener: Dibyendu Nandi)

**16:45 – 17:15 Tea/Coffee Break**

**17:15 – 18.15: Closing Session: Felicitating Arnab Rai Choudhuri**
*Lecture by Arnab Rai Choudhuri:  "My Life and my Journey through Solar Physics"*
*Felicitation Ceremony*

**19:30 – 21:30: Dinner / Opening Reception for IAUS 340 (on-siteW)**

---

The programme of the workshop.  Vasilis Archontis's talk was cancelled, as he could not come. All the other talks were given.

# My Life and My Journey through Solar Physics

When Dibyendu started inviting several very busy persons to give talks at this workshop, I had feared that he would receive many polite regret letters. I know that such workshops are sometimes organized for very eminent scientists. But, when such a workshop is organized for an average scientist like me, I wondered who would want to attend it. I am really overwhelmed that so many of you are here and so many outstanding scientists have agreed to speak on this occasion.  This only shows that I have received a kind of love and affection from the community which is completely disproportionate with my modest scientific achievements.

Dibyendu gave me two options for my talk: I could make some observations on the present and the future of solar physics, or I could talk about my life and my association with solar physics.  Now, to make some observations on the present and the future of solar physics, one needs to have lots of wisdom, which I lack.  On the other hand, I thought that even a fool can speak about his life and career. However, after choosing this second option, when I started preparing the talk, I realized how hard it is! I have given many scientific talks and I know how to prepare such a talk.  But I have never given a talk like this one and I am completely unable to judge whether the talk I have prepared will be at least moderately interesting to the audience or totally boring. So, you are going to be my guinea pigs on whom I shall try out this talk, although this is an experiment which will never be repeated – I cannot turn 60 twice!

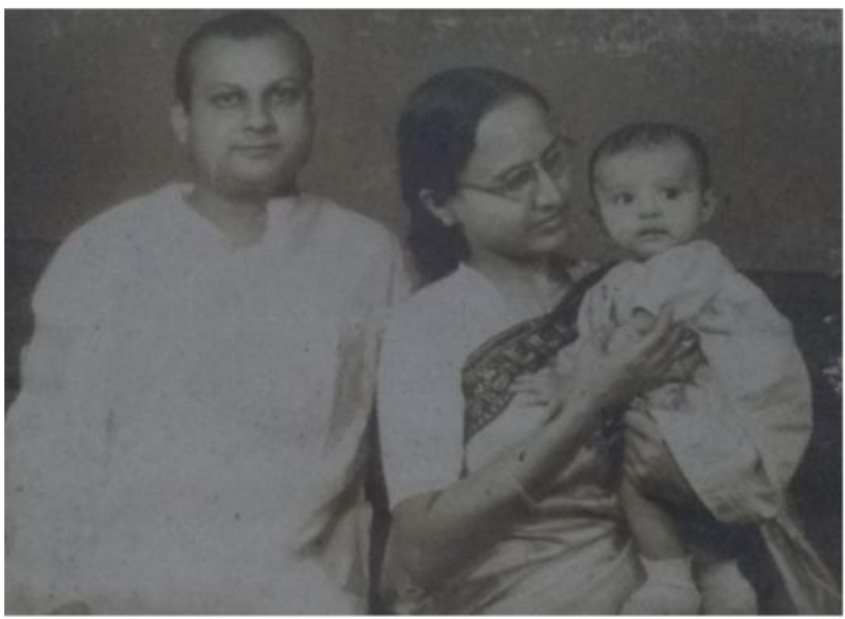

Figure 1. With my parents at the age of a few months.

I was born in the bustling city of Kolkata a few years after Indian independence.  Since India gained independence from the British rulers at the stroke of a midnight, people belonging to the generation born around India's independence are often called midnight's children – a phrase coined by novelist Salman Rushdie. I am one of midnight's children. Nowadays, when a young couple gets a new baby, hundreds of photographs get uploaded in the social media in the next few hours. That was not the case 60 years ago. Figure 1 is probably the first photograph of mine taken in a studio when I was a few months old.

Let me say a few words about my parents. Without their tremendous support for my academic career, I could not become what I am today. They were classmates in Presidency College – arguably the most important college of India for more than a century. Starting from physicists Saha of Saha ionization equation fame and Bose of Bose-Einstein statistics fame, to film director Satyajit Ray and Nobel-winning economist Amartya Sen – virtually all important persons of Kolkata had studied in that college. Till the early 1940s, it was a boys' college. My mother was a student in the second batch after the college became co-educational. The marriage of my parents was the first marriage between classmates in the history of Presidency College. Since I am their first child, I have some fame in certain Kolkata academic circles as being the first child of a marriage between classsmates in Presidency College. Normally you have to work very hard to be the first in something important. I am lucky that I could be the first in something merely by being born!

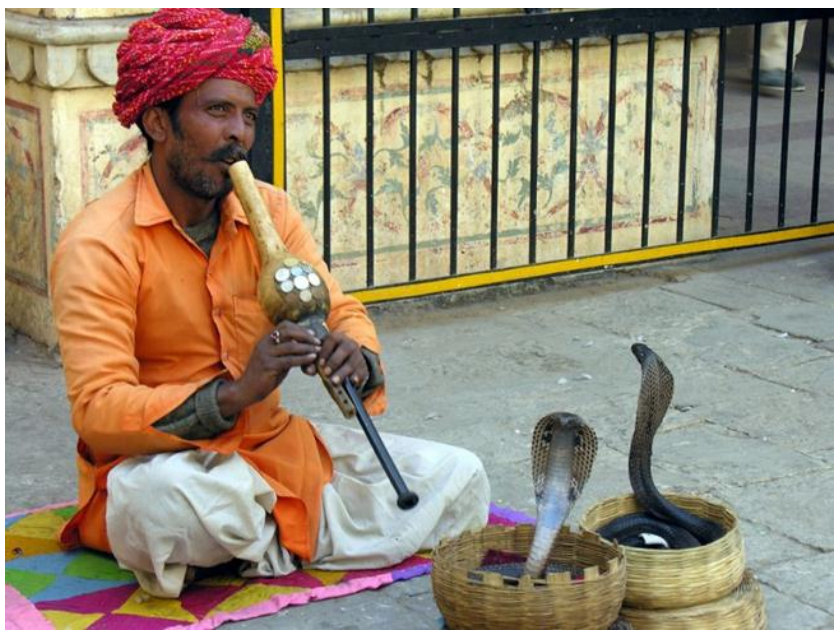

Figure 2. A typical snake-charmer on the footpath of an Indian city.

India of my childhood days was very different from India of today. Two centuries of colonial rule had completely impoverished the country. Shortly before the Indian independence, there was a terrible famine in eastern India in which three million people are estimated to have perished – many of them dying of starvation in the streets of Kolkata. Even during my childhood days, if you stood at a street corner of Kolkata, you would see many people who looked visibly poor and malnourished. There would be many beggars around. Many persons would be walking barefoot, as they could not afford to buy shoes. But, in spite of this poverty, our childhood was much more colourful than the childhood of the next generation. In those days, the city and the village intermingled in Kolkata. There were some people raising cows behind our house. These cows would often stroll in front of our door. Then there were people of many strange professions, such as travelling salesmen who would carry their fares in baskets on their heads and cry out describing their fares. I used to be especially fascinated by snake-charmers, who would make snakes dance by playing the flute (Figure 2). I remember being heart-broken when snake-charmers disappeared from the streets of Kolkata a few years later. At that young age, I did not realize that it was a sign of economic prosperity. Think of the abject poverty that would force a man to risk his life catching poisonous snakes in forests and then to make them dance in the streets of Kolkata for a pittance. In those days, most people dressed traditionally. Many of our teachers in the school and the college wore the beautiful gentlemen's dress dhoti-panjabi, which you can see my father wearing in Figure 1. In a few years, the dhoti disappeared with the disappearance of another profession like the snake-charmer – the family washerman. In my childhood days, each family would have a washerman who typically would come once a week to collect the dirty clothes and then would wash, starch and iron them. Once the family washerman was gone, it became almost impossible to maintain the dhoti.

I did my bachelor's degree in Presidency College where my parents had studied. There I had an extraordinary teacher Amal Raychaudhuri (Figure 3), whose physics teaching was so powerful that it could transform people's lives. When Raychaudhuri was a student, nobody in Kolkata knew much about general relativity. He learnt it on his own and formulated what is now called the Raychaudhuri equation. It is the starting point for proving the singularity theorems of Hawking and Penrose. Many years later, when I myself had to teach students, I had Amal Raychaudhuri as my role model and consciously tried to imitate his teaching style. Even now, when I have to teach a difficult subject, I try to imagine how Amal Raychaudhuri might have taught it.

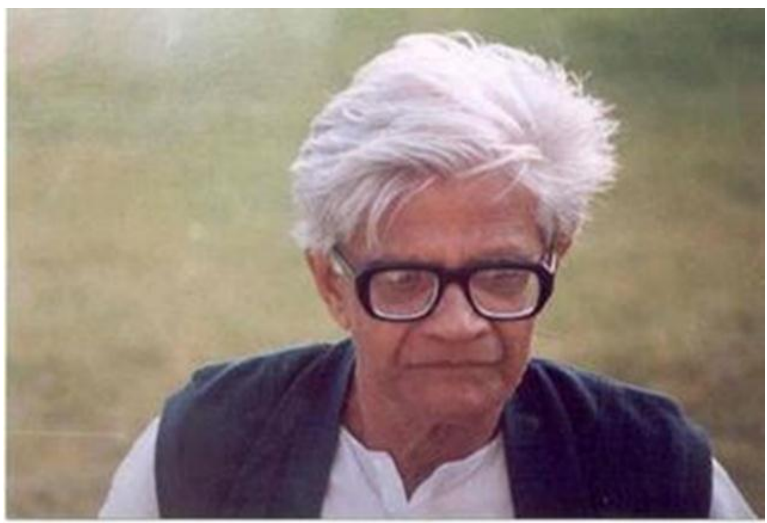

Figure 3. Professor Amal Raychaudhuri of Presidency College.

My next stop for master's degree was IIT Kanpur. That was a place where some of the top students from every batch would go for graduate school in American universities. Almost as soon as we joined, a peer pressure started building up on us for taking GRE, TOEFL, etc. Through that process, I ended up at the University of Chicago as graduate student in 1980.

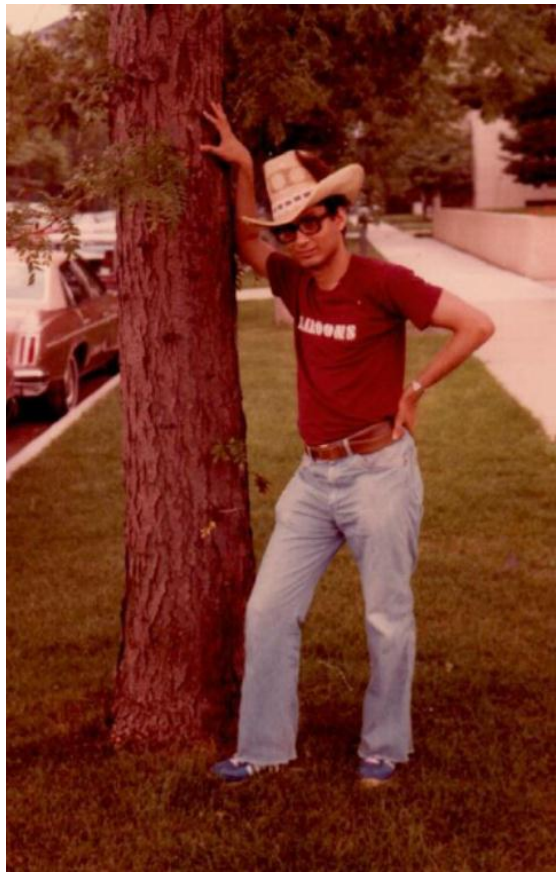

Figure 4.  As a graduate student at the University of Chicago Campus.

If you want to know what I looked like in those days, Figure 4 shows me on the campus of the University of Chicago. There I had the rare good luck of having an extraordinary scientist and extraordinary human being as my PhD supervisor: Gene Parker (Figure 5).  If I start talking about Gene, I can talk for hours.  So, I would rather refer you to my popular science book *Nature's Third Cycle*.  There I have reminisced about my Chicago days and about my association with Gene. Read that book to find out what it was like to work under the supervision of the world's greatest solar physicist. We worked in the building Laboratory for Astrophysics and Space Research.  In those days, most of the astrophysicists at the University of Chicago were in a building called Astronomy and Astrophysics Center. The building Laboratory for Astrophysics and Space Research (Figure 6) mainly housed the large cosmic ray research group. However, two of the most beautiful offices in that building were given to two theoretical astrophysicists – Chandrasekhar and Gene Parker. The upper right corner room seen in Figure 6 was Chandrasekhar's office and the upper left corner room was Parker's office. My office was in the room next to Parker's office. Chandrasekhar had stopped taking students after a heart attack in the 1970s and there were no students working with him when I was in Chicago. For a while, I was the only theory student having a very nice office in this building. When I was attending the first AAS meeting of my life, somebody asked me at the dinner table, "How big is your theory group?" I replied: "It is a very small theory group with only three members." The next question was, "Who are the members?" I casually said: "Oh, besides myself, the other two members of our small theory group are Subrahmanyan Chandrasekhar and Eugene Parker."

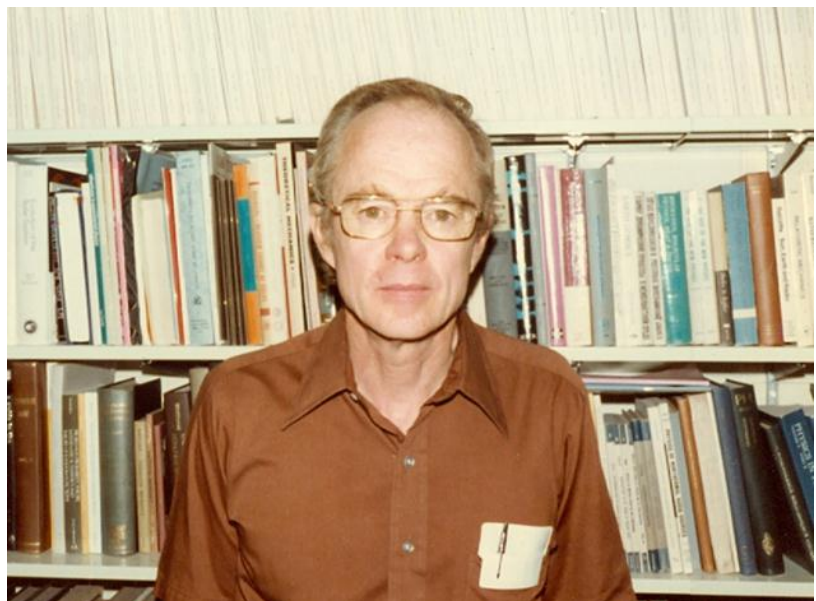

Figure 5. Professor E.N. Parker in his office at the University of Chicago.

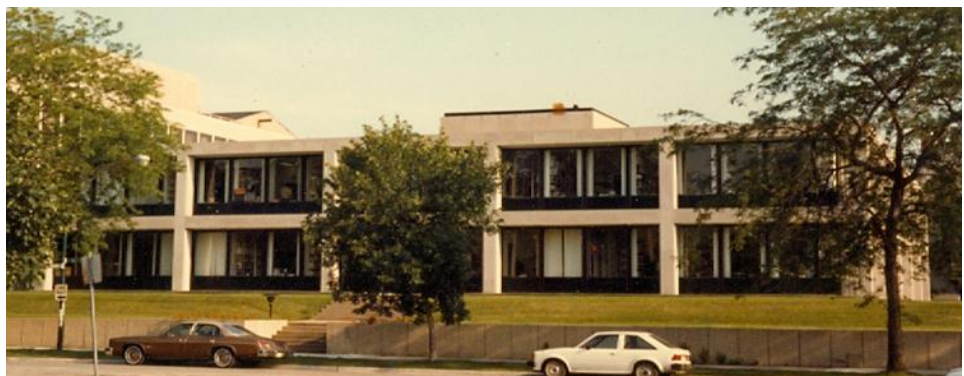

Figure 6.  Laboratory for Astrophysics and Space Research.

After obtaining my PhD from Chicago in 1985, I was a postdoc for two years at High Altitude Observatory (HAO) in Boulder.  In those days, HAO was located inside the striking building designed by the legendary architect I.M. Pei in the foothills of the Rocky Mountains (Figure 7).  An HAO postdoc had the wonderful freedom of working with anybody in HAO and was not bound to any particular scientist.  As many of you surely know, Gene Parker never touched a computer.  All his calculations were done with paper and pencil.  So, I had no opportunity of learning to do numerical simulations when in Chicago. I thought that I can devote my two years at HAO profitably for learning to do numerical simulations. Peter Gilman (Figure 8) at HAO had done some impressive simulations about solar convection and the dynamo process. I asked him if I could work with him and learn to do numerical simulations. Several persons in the audience probably know that years later I was involved in an unpleasant controversy with Peter. You can read an account of this controversy – at least my version of it – in my book *Nature's Third Cycle*. Today I shall not talk about this controversy. Working with Peter, I got involved in a research project which virtually determined my subsequent career path. Since I consider the Choudhuri & Gilman (1987) paper to be the decisive paper that shaped my career, let me say a few words about it.

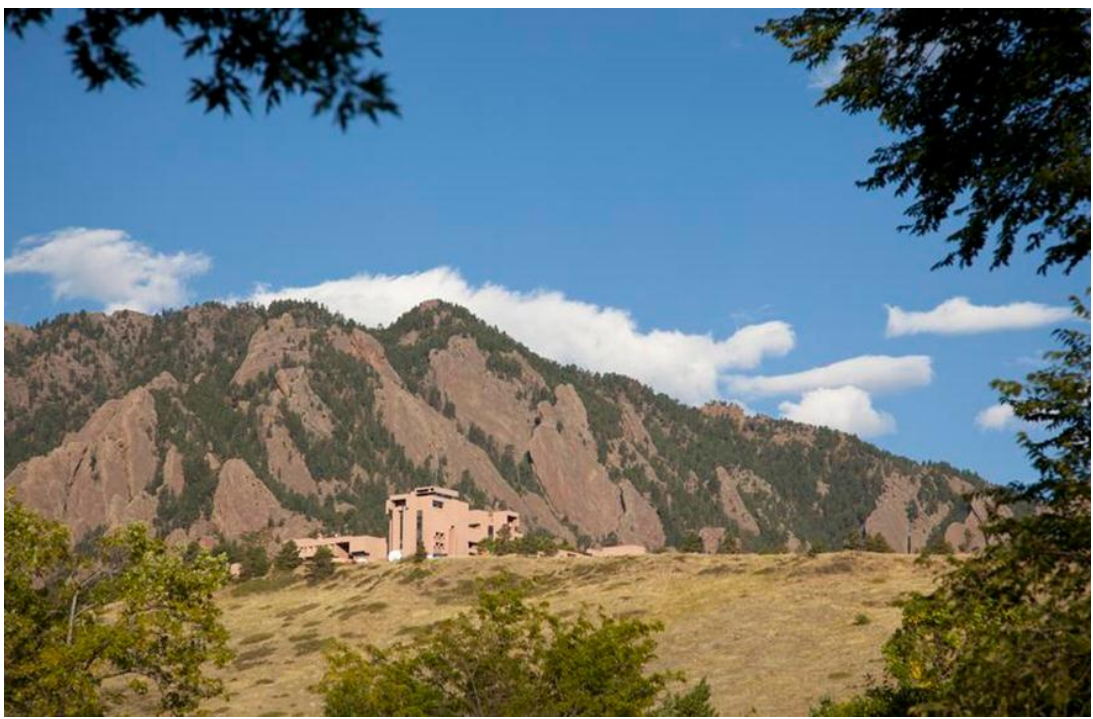

Figure 7. The Mesa Laboratory in Boulder which housed HAO.

We normally do not see sunspots very near the solar equator. Peter had the idea that, when flux tubes rise due to magnetic buoyancy near the equator, probably they get deflected a little bit by the Coriolis force. Since nobody had looked at the effect of the Coriolis force on rising flux tubes and this work had to be done numerically, Peter suggested that I could learn numerical simulations by working on this problem. This seemed like a fairly routine research problem and, if we had got the results we expected, it would lead to an average routine paper. To my utter surprise and horror, I discovered that the Coriolis force was much more important in this problem than what anybody suspected before and the solar dynamo models available at that time could possibly not be correct. The venerable α-effect – the centrepiece of dynamo theory – may not be operative in the Sun. You may wonder what emotions such a discovery evokes. Is the discoverer overjoyed? I can say that at that time it was a feeling of sheer terror for me. We who work on theoretical physics often tend to regard our equations like children's toys with which we play our little games. Only rarely we become aware that these equations have a hidden power and may tell us things we don't want to hear. Normally only the greatest of theoretical physicists working on truly important problems occasionally have a glimpse of this mystifying power of mathematical equations. I am lucky that I, an average physicist working on what seemed like a routine research problem, suddenly had a glimpse of this terrifying and mysterious power of mathematical physics. This unexpected discovery kept me thinking for years how to resolve these difficulties – ultimately leading to the formulation of the flux transport dynamo model. After that, I spent much of my career working out its consequences.

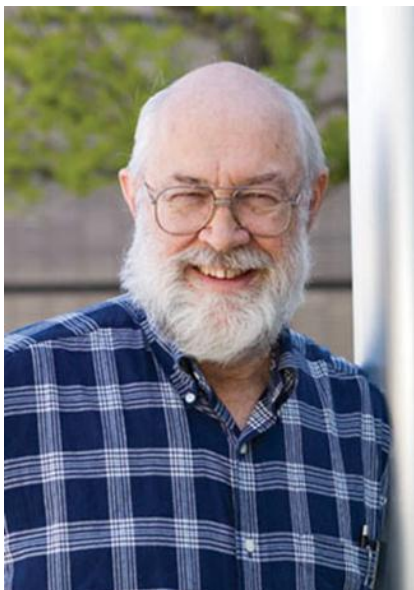

Figure 8. Dr. Peter Gilman of HAO.

I had to decide what I wanted to do after my postdoc at HAO. Some of us midnight's children had grown up with the idealism that we had to build our newly independent nation. I started looking for a job in India. One senior scientist at HAO was B.C. Low. He grew up in Singapore and did his PhD under Gene Parker's supervision about a decade earlier than me. He described to me his own experience with Singapore and told me: "Things are going well with you in US. You should not have difficulty in getting a good job in US. If you go back to India, that will be the end of your scientific career." I was lucky to receive job offers from both of India's topmost academic organizations – Indian Institute of Science (IISc) in Bangalore and Tata Institute of Fundamental Research (TIFR) in Mumbai. In many countries, you will find that their two top academic organizations would be very similar. Oxford and Cambridge in England are not too different from each other. But these two top organizations in India are as different from each other as two academic organizations can be (Figure 9)! So, it was not easy to select between the two. One thing about TIFR which repelled me was that I would have to stay inside a matchbox-type air-conditioned building from the morning to the evening if I worked there. On the other hand, IISc has a verdant green campus with various departments scattered around that campus. However, what finally made me take a decision in favour of IISc was the Joint Astronomy Programme or JAP. Let me say a few words about it.

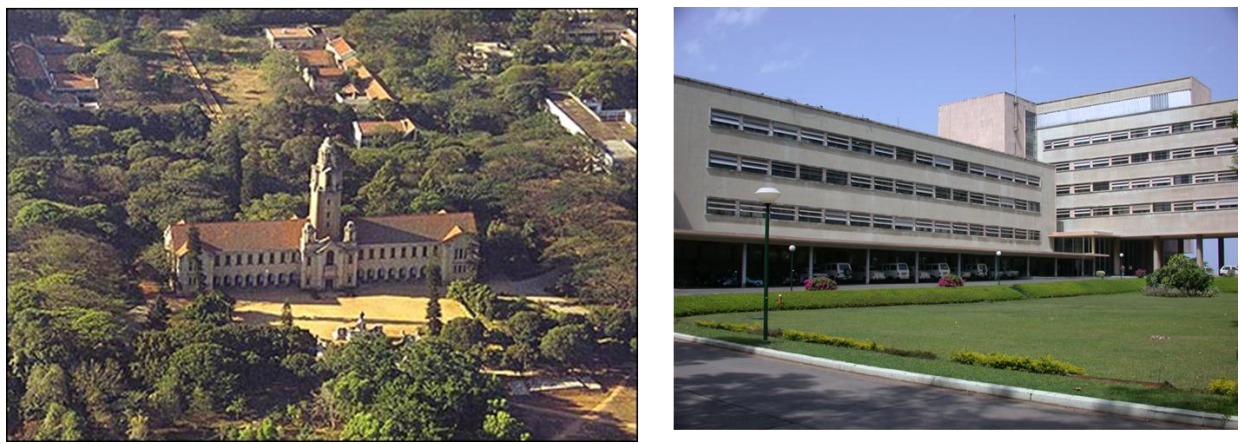

Figure 9. Indian Institute of Science (IISc) and Tata Institute of Fundemantal Research (TIFR).

In the 1920s, M.N. Saha working in India formulated the Saha ionization equation and developed his famous theory of stellar spectra. However, there was no continuity of astrophysics research in India after that. There was a complete break and, at the time of Indian independence, there was no research on modern mainstream astrophysics being done anywhere in India. In the 1960s and 1970s, a few Indian astrophysicists trained abroad returned to India and started the first astrophysics research groups in independent India. The three gentlemen whose photographs you see in Figure 10 were among those pioneers who initiated modern astrophysics in India. Unfortunately Vainu Bappu passed away at a relatively young age and I never met him. All the other pioneers of modern astrophysics in India were still in service when I returned to India in 1987 and I knew all of them personally. Working on astrophysics in India at that time, one felt that one was in a young field. There were no retired astrophysicists anywhere in India at that time! In the late 1980s, the number of people in India who would publish papers in ApJ, A&A or MNRAS with reasonable regularity was quite small – definitely not larger than 25 or 30. Till the early 1980s, no place in India offered courses on astrophysics. Any student who wanted to do PhD in astrophysics had to learn the subject on his or her own. As astrophysics started becoming a more mature field, there was a crying need for a proper graduate programme.

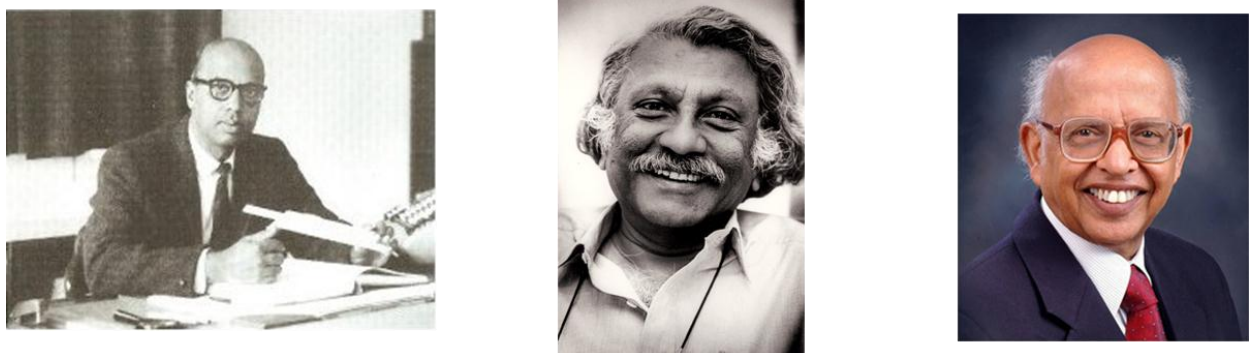

Figure 10. M.K. Vainu Bappu, V. Radhakrishnan and Govind Swarup.

In the 1980s Bangalore was the city in India with the largest number of astrophysicists working basically in three groups – at Indian Institute of Astrophysics (IIA) headed by Vainu Bappu, at Raman Research Institute (RRI) headed by V. Radhakrishnan and in the TIFR radio astronomy group headed by Govind Swarup. Although the main campus of TIFR is in Mumbai, the radio astronomy group was in Bangalore, from where they could operate the amazing Ooty Radio Telescope they had built. These three visionaries – Bappu, Radhakrishnan and Swarup – initiated Joint Astronomy Programme or JAP in 1982. Although IISc had no astrophysicists on its faculty, it was chosen as the nodal institute because it was a large organization which could grant PhD degrees. Students selected in JAP would undergo course work at IISc for a year taught by scientists of all the collaborating astrophysics groups and afterwards a student could work for PhD with a scientist in any of these groups. The Programme started with great enthusiasm. Figure 11 shows photographs of Rajaram and Srini, who were legendary teachers of JAP in its early years. However, in those days before e-mail when telephones were virtually dysfunctional in India, it required Herculean efforts for scientists of other organizations to do even simple things like scheduling a course at IISc or carrying on the various official formalities of the students. After the initial enthusiasm evaporated in 2-3 years, it became very hard to manage JAP. At that time C.N.R. Rao, the most distinguished living Indian scientist today, was the Director of IISc. He decided to hire two young faculty members at IISc to manage JAP. Chanda Jog (Figure 11) joined in one position. I joined in the other position a few months later.

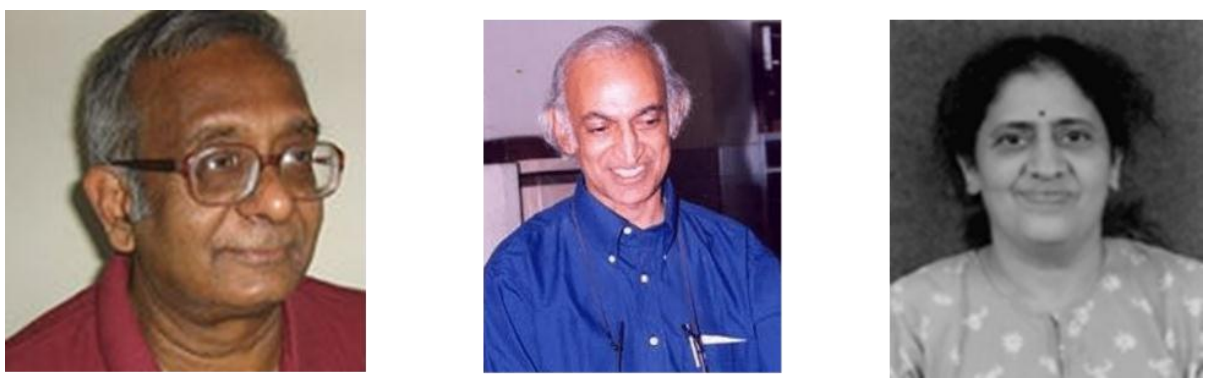

Figure 11. Rajaram Nityananda, G. Srinivasan and Chanda Jog.

I returned to India in 1987 after spending seven years in the US. When I was applying for jobs in India, I had come to India for a visit and spent a few days in Bangalore. One grandfatherly gentleman with whom I had a long conversation was Prof. S. Ramaseshan, who was the Director of IISc when JAP was initiated. He told me: "We started JAP with lots of hope, but it has run into rough waters. Now my successor C.N.R. Rao is trying to save the Programme by hiring two lecturers to run it. Don't take up one of these positions. These two poor lecturers, who would have no previous experience with the system, will have to do many thankless chores and will not have time for anything else. In any case, IISc has no astrophysics research facilities. Keep away from IISc." Even after I joined IISc, I kept receiving cautionary advice. One departmental colleague a few years older than me was Arup Raychaudhuri, a brilliant experimental condensed matter physicist. He told me: "You have made a big mistake by joining IISc. As soon as possible, shift to a place which has a good astrophysics group. After all, we know that JAP is a dying programme. At this stage of your career, if people identify you with a dying programme, that will not be good for your career." Well, I have to keep you till midnight if I want to tell you all that some of us had to do in order to transform this dying programme into a programme which has trained probably more than 60% of India's active astrophysicists today. Let me tell you only this. I know that my papers and books are appreciated by many. But, if I am asked what in my view is my greatest academic contribution, I would say that I have played some role – perhaps the decisive role – in changing the destiny of India's first and most important graduate school in astrophysics when I was barely in my early thirties.

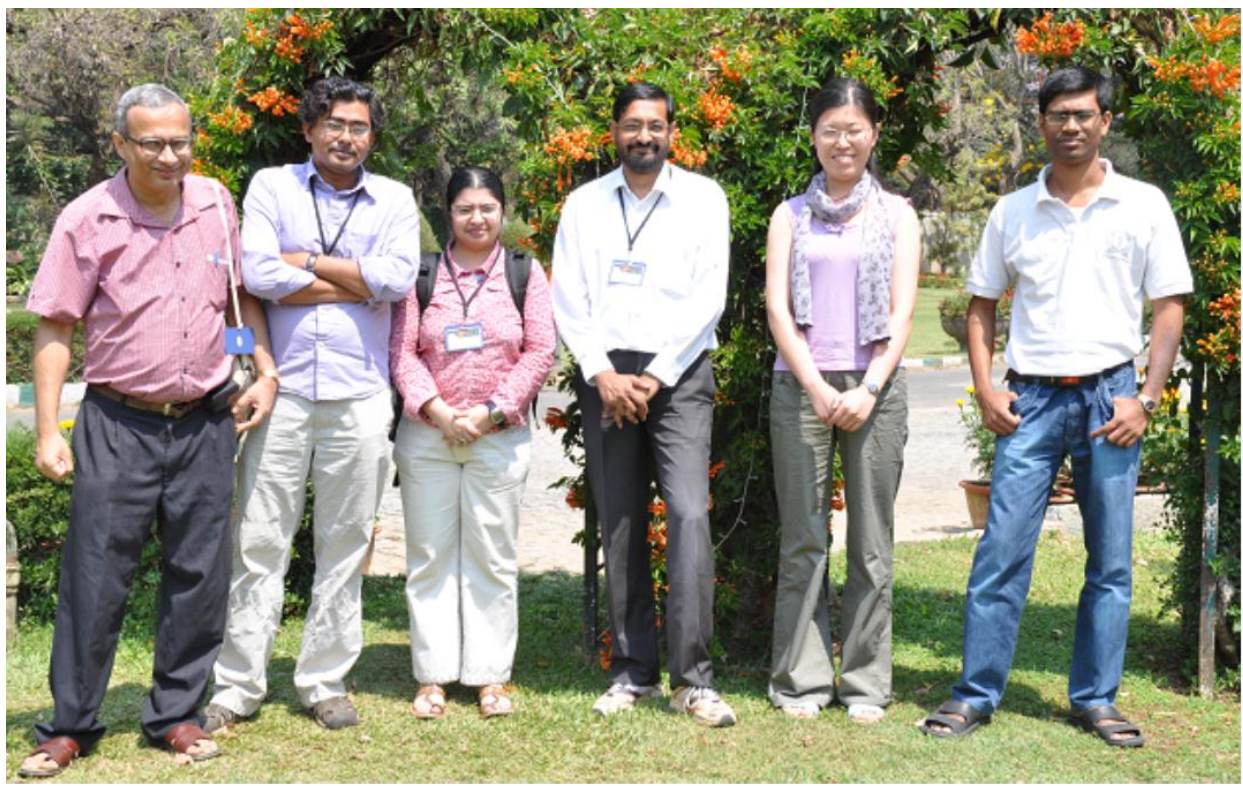

Figure 12. With several of my students: I, Dibyendu Nandy, Piyali Chatterjee, Dipankar Banerjee, Jie Jiang, Bidya Karak.

At IISc I had the rare good fortune of attracting exceptional PhD students one after another. Whatever little I have achieved in the last 30 years has been due to them: Sydney D'Silva, Mausumi Dikpati, Dibyendu Nandy, Piyali Chatterjee, Jie Jiang, Bidya Karak, Gopal Hazra. Many of them were JAP students and others also had gone through the JAP course work. I, of course, shall not forget my Chinese student Jie Jiang. Again, read *Nature's Third Cycle* to find out the circumstances under which she worked for her PhD with me. Without exception, each of my students has done at least one famous work during PhD which is regarded as a classic in solar physics. There are many solar physicists around the world who would be able to describe the PhD thesis works of all of my students. I am often asked: How I managed to supervise such exceptional PhD theses one after another? What is my secret? I have decided to reveal the secret today. There are many mid-career scientists in the audience who supervise students for PhD. They may have expectations that, after I reveal my secret, they may try it out on their students. But, after I reveal the secret, you will realize that probably nobody else can supervise students the way I had done! Nowadays, to do good research in solar physics, you need to have certain technical skills. Those who have interacted with me closely know that I am completely deficient in all technical skills. I cannot do challenging analytical calculations. I cannot write difficult big codes. I cannot do complicated data analysis. After students decide to work with me – either because I am a charismatic teacher or for any reason whatsoever – within a few months they realize that I totally lack all technical skills. Their first reaction after finding this out is usually of surprise and disbelief. Although all my students have been very polite, I could read their minds. They would start thinking: "My God! I have got a supervisor who does not seem to know any techniques for doing research. What will happen to me? How shall I get my PhD?" Eventually, out of desperation, my students would start learning the technical skills on their own. In that process, they would become completely independent and do important work. I am sure that all the mid-career scientists present in the audience have many technical skills. Unless you are totally deficient in all technical skills, you cannot inspire the feeling of total desperation I could inspire in my students and you cannot supervise students like me!

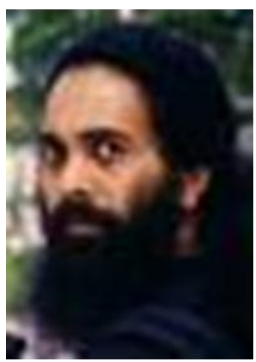 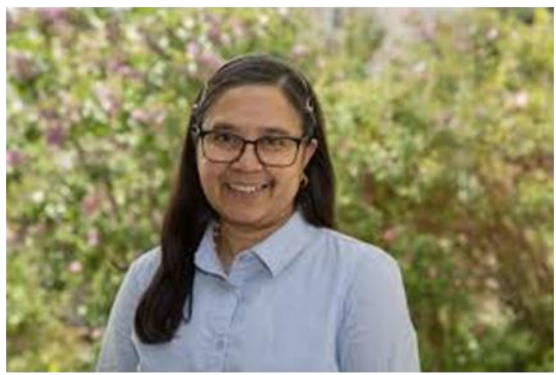 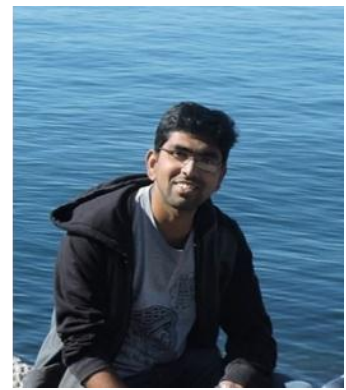

Figure 13. My students not present in the group photograph: Sydney D'Silva, Mausumi Dikpati, Gopal Hazra.

I may also point out that there are certain hazards of supervising very bright students. My present student Gopal is my only student who has never scolded me. I have received severe scoldings from all my other students on various occasions. Some of you who have known many of my students well can probably guess who was my student from whom I received the maximum amount of scolding: Dibyendu Nandy. Let me tell you how our paper which was published in *Science* got written. After Dibyendu got some interesting results, he told me that we should try to publish these results in *Nature* or *Science*. Till that time, I have never tried to publish anything in one of those so-called high-impact journals. I suggested that we might write this paper also for a regular journal like ApJ or A&A. Dibyendu told me: "You have got some serious attitudinal problems, which you have to get over. You have done some good work on dynamo theory. But nobody gives you much importance. Once you publish something in *Nature* or *Science*, you will find that people will give you much more importance." I have known of supervisors who had told their students: "You have done some good work, but people do not give you enough importance because of your attitudinal problems." I am not aware of any other example of a student telling that to his supervisor! Then Dibyendu made xerox copies of about ten papers on theoretical solar physics which had appeared in *Nature* or *Science* in the past few years (in those days one had to go to the library to look up papers and could not download them from internet!) . Dibyendu placed those papers on my desk and said: "A paper for *Nature* or *Science* has to be written very carefully. You have good writing skills: you have written a book. If you read these papers, then you will have an idea how to compose a paper for *Nature* or *Science*. After that, prepare the paper based on our results." I was extremely busy in the next few days and could not work on this. When Dibyendu enquired a few days later and found that there was no progress, he lost his temper: "You always say that you are very busy. But I never see you doing anything really important. You just sit in your room and only read various things or write e-mails. Anyway, I give you one more week. You better prepare the first draft of our paper within a week." I was much relieved when Dibyendu finished his PhD and left!

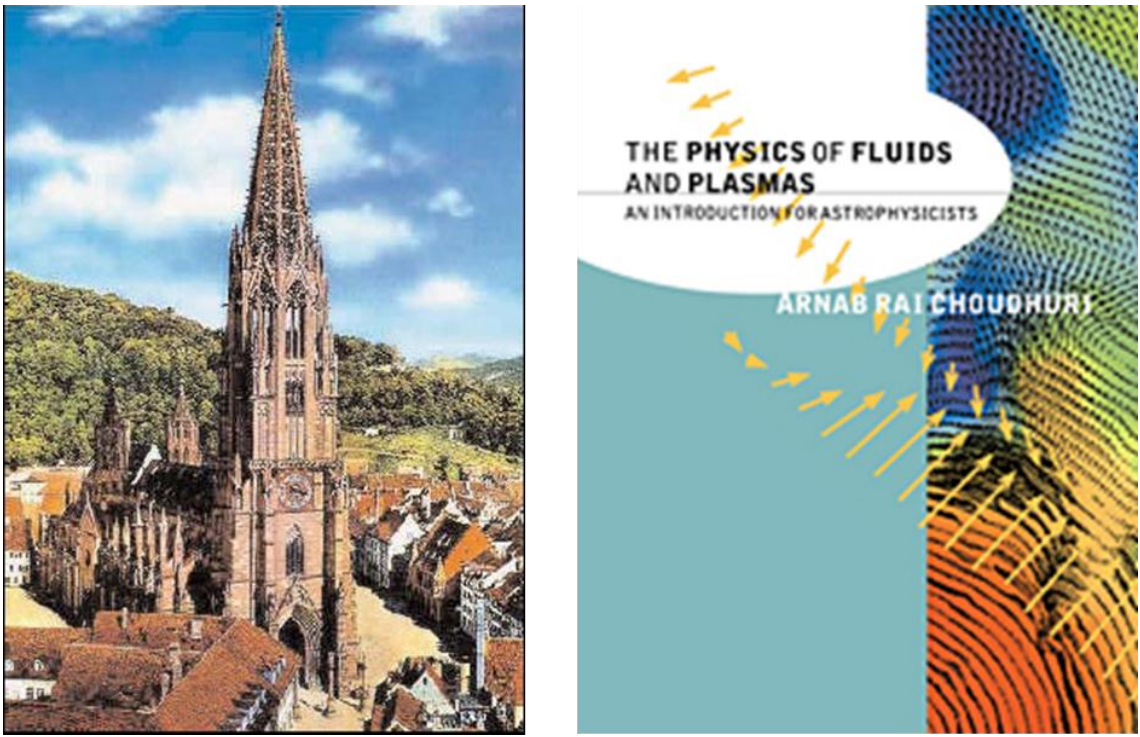

Figure 14. The cathedral in Freiburg and my first book which I began in Freiburg.

I should say a few words about the wonderful sabbatical year I spent at Kiepenheuer-Institut in the beautiful city of Freiburg built around a magnificent medieval cathedral at the edge of the Black Forest (Figure 14). In some ways, that sabbatical year was probably the most creative year of my life. During that year, I wrote the major part of my first and most important book *The Physics of Fluids and Plasmas*. I know that my other textbook *Astrophysics for Physicists* sells many more copies every year than this book. But, about this first book of mine, I would like to echo what Walt Whitman said of his *Leaves of Grass*: "This is no book. Who touches this, touches a man." Being busy with writing this book, I managed to get time to write only one short paper during this sabbatical year, based on my interactions with Manfred Schüssler, who was at Kiepenheuer-Institut at that time. That short paper is presumably the most important paper of my life. That paper is generally regarded as the foundational paper of the flux transport dynamo model, the currently favoured model of the 11-year sunspot cycle. If you read the 4-page Choudhuri, Schüssler & Dikpati (1995) paper now, you will think that it is a very straightforward paper, putting forth some rather simple and obvious ideas. In science, coming up with simple ideas is often the most difficult task. I myself have difficulty in believing now that it required Herculean efforts to come up with those simple ideas at that time

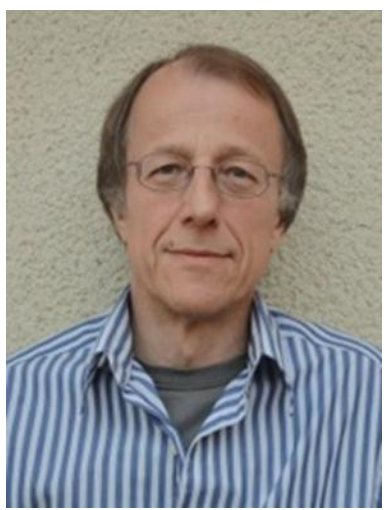

Figure 15. Professor Manfred Schüssler.

Let me now talk about an initiative of mine that I am particularly proud of. When I returned to India, outside of Japan there was very little good solar physics research done anywhere else in Asia. We looked towards the West for appreciation, collaboration, visiting positions, etc. As solar physics communities grew in several Asian countries, I started arguing that there should be more interactions among the solar physics communities of the Asia-Pacific region. On 24 May 2010, I wrote an e-mail to several senior solar physicists of this region – perhaps the most important e-mail I ever wrote in my life – suggesting that we organize Asia-Pacific Solar Physics Meetings (Figure 16). So far, four Asia-Pacific Solar Physics Meetings (APSPM) have taken place in Bangalore, Hangzhou, Seoul and Kyoto, which proved successful beyond my wildest expectations. It is now recognized as a very important meeting, and being asked to give an invited talk in one of these meetings is already considered very prestigious. I have never been asked to give an invited talk in these meetings. I guess, one has to be very good to receive such an invitation!

I also strongly felt that there should be a good astrophysics journal from the Asia-Pacific region. I have been very much involved, along with Jingxiu Wang, in establishing the journal

*Research in Astronomy and Astrophysics* (Figure 17). However, I have to confess that this effort has not been so successful like the Asia-Pacific Solar Physics Meetings. Perhaps time has not come yet for a strong astrophysics journal from our region. But I am confident that there will be such a journal one day – perhaps within the active career of some of the younger participants of this workshop.

> Date: Mon, 24 May 2010 10:21:22 +0530 (IST)
> From: Arnab Rai Choudhuri <arnab@physics.iisc.ernet.in>
> To: wangjx@bao.ac.cn, hzhang@bao.ac.cn, fangc@nju.edu.cn, moonyj@khu.ac.kr, Jong chul Chae <chae@astro.snu.ac.kr>, chou@phys.nthu.edu.tw, shibata@kwasan.kyoto-u.ac.jp, sakurai@solar.mtk.nao.ac.jp, saku.tsuneta@nao.ac.jp, kit@iszf.irk.ru, paul.cally@sci.monash.edu.au, cairns@physics.usyd.edu.au
> Cc: hasan@iiap.res.in, P Venkatakrishnan <pvk@prl.res.in>
> Subject: Asia-Pacific solar physics meeting
>
> Dear fellow solar physicists,
>
> Some of us solar physicists in India have been wondering if it will be worthwhile to organize an Asia-Pacific meeting on solar physics. This is to share with you our ideas on this matter and to receive your feedback.
>
> The Asia-Pacific region is arguably the region in the world where solar physics is growing at the fastest rate. While science is an international venture, regional meetings do serve important purposes. American solar physicists meet in the SPD meetings. European solar physicists also have their own meetings. The main aim of an Asia-Pacific meeting will be to foster cooperation and collaboration within our region. We shall certainly not think of excluding anybody from such a meeting. Any solar physicist from anywhere in the world will be most welcome to attend this meeting. We may even ask a few solar physicists from outside this region to give invited talks. As for the scope of the meeting, in addition to the traditional topics in solar physics, we can cover space weather, heliospheric physics and solar-stellar connection.

Figure 16. The beginning of my e-mail proposing APSPM.

Before I end, I want to say a few things about my personal life. I got married to Mahua on an astrologically inauspicious day (Figure 18). At the beginning of every Indian year, astrologers declare certain days to be auspicious days for marriage. As a protest against astrology, I insisted that I would get married only on a day that astrologers considered bad. I, of course, do not know whether our married life would have been happier if we were married on an astrologically auspicious day! Mahua is a physics teacher at a reputed women's college in Bangalore. I could devote so much time to science only because Mahua shouldered a grossly disproportionate amount of family responsibilities. I have continuously felt guilty for not giving enough time for the family. If I say that Mahua always provided me ungrudging support, then I may be distorting the truth a little bit. But she tolerated me as a husband over the years, which not many women could do! Figure 19 shows our two sons, whose childhood dreams were to become footballers. Now both of them are studying for academic careers.

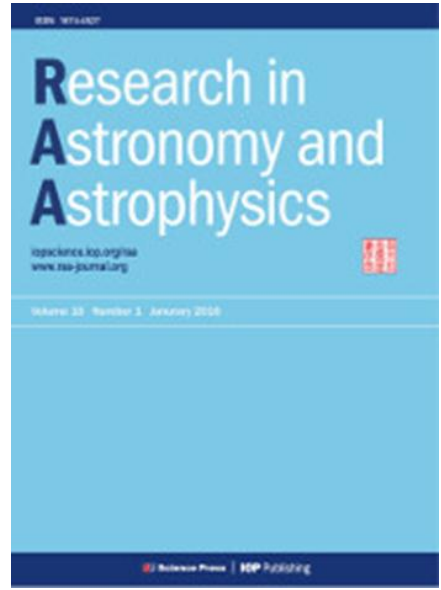

Figure 17. The astrophysics journal from the Asia-Pacific region.

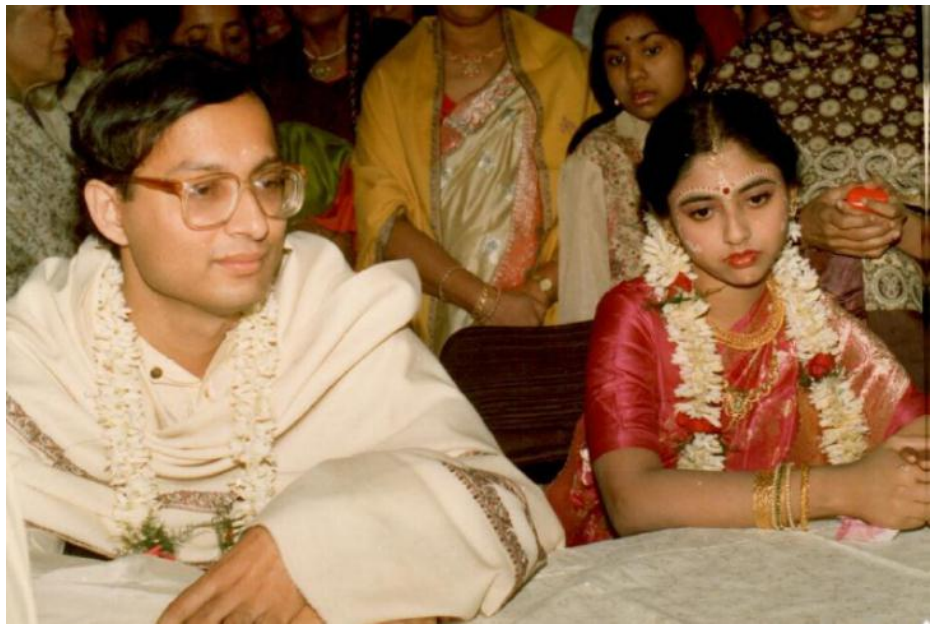

Figure 18. With Mahua on our marriage day.

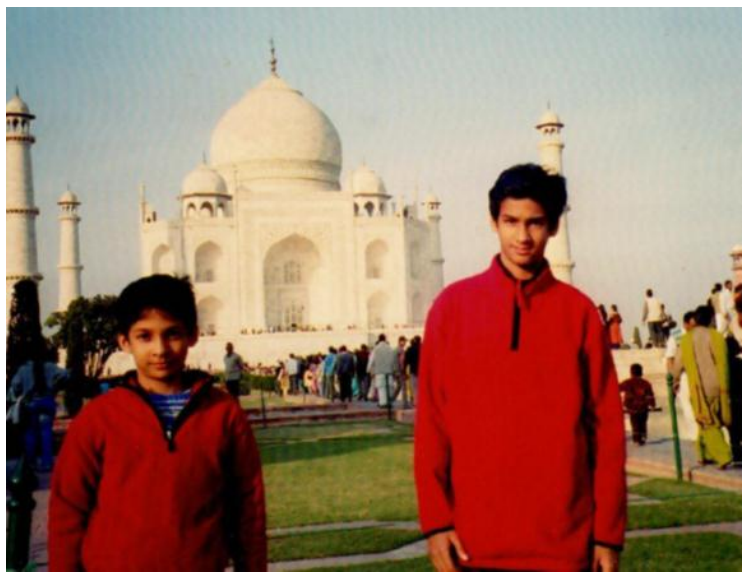

Figure 19.  Our two sons Mukul and Arka.

About four years ago, I was diagnosed for cancer.  Initially doctors suspected it to be lung cancer and I had an interesting week when I was under the impression that I had about 3-4 more months to live.  A difficult biopsy eventually established that what I had was Hodgkin's lymphoma – a cancer of the lymph node – usually curable with modern medicine.  I had to undergo many chemotherapies and radio therapies for about eight months. For nearly a year, it was almost impossible for me to do any serious work.  My last student Gopal had started working with me a few months before this period.  Although I was not in a position to provide any research supervision for a long time, Gopal refused to look for an alternative supervisor.

Since my health was unsteady for a while even after the treatment and there was a fear of the relapse of cancer, I felt that it would not be correct for me to take any more students.  This also meant that I could not carry on the research the way I had been doing.  I always had an interest in history of science.  At that stage, I took the decision that I would phase out my research in solar physics after Gopal graduates and devote myself completely to history of science.  I am now tremendously excited about my research in this new field and would love to tell you what I am doing in history of science.  But I refrain from doing so due to the lateness of the hour and because it may not interest all of you.  So, let me just show you a photograph (Figure 20) I feel particularly proud of.  The Royal Society of London has one of the world's most magnificent archives of scientific materials.  When I visited these archives a few months ago, I had the rare honour that the archives staff offered to take my photograph with the most prized possession of the Royal Society – the manuscript of the greatest physics book ever written. I am not sure if you can discern the title of the manuscript in Figure 20. But you all know what the greatest physics book of all time is! I had held the manuscript of Newton's *Principia* with these hands of mine.

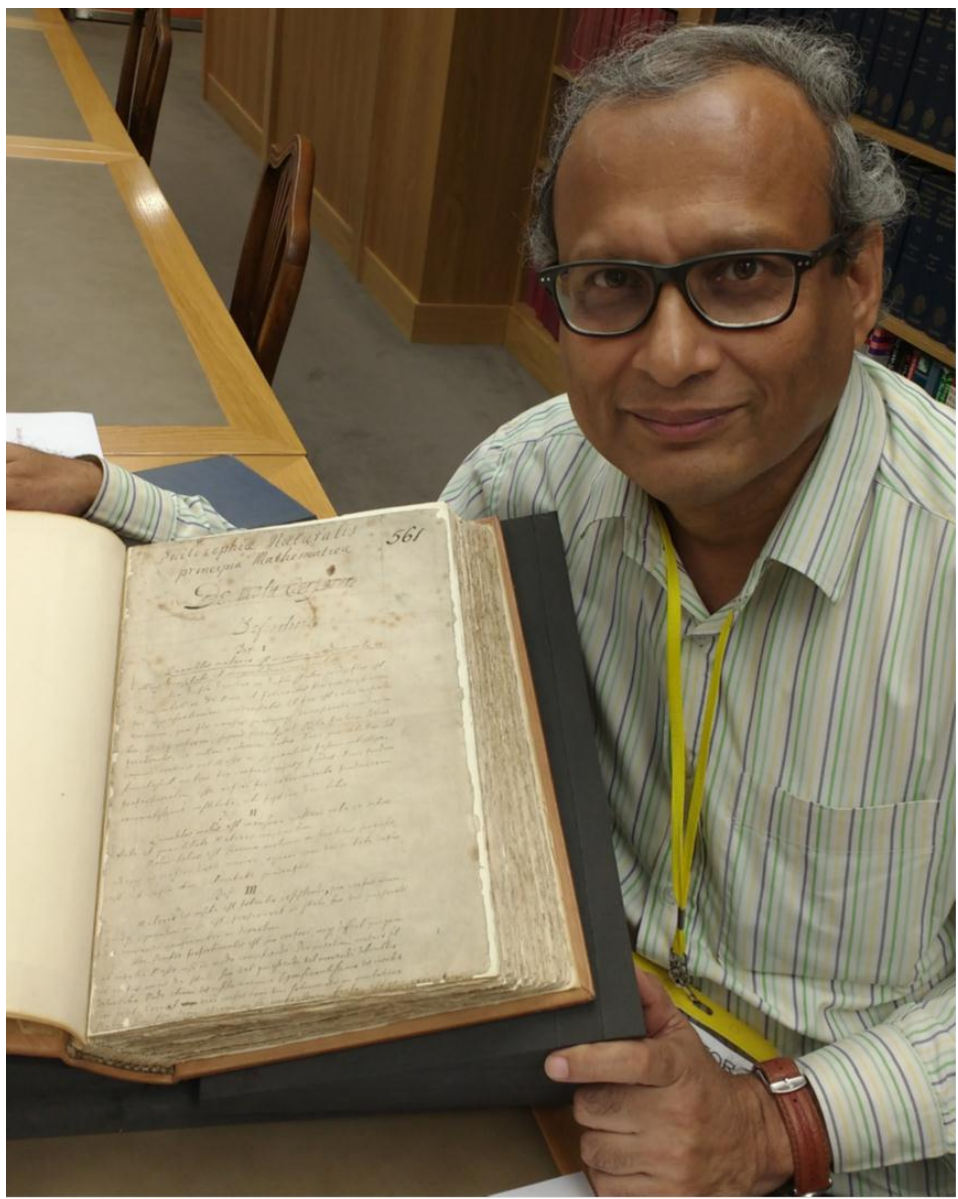

Figure 20. Holding the manuscript of Newton's *Principia*.

One of the greatest rewards of working in a scientific field is that you make so many friends. I have been meeting many of you in many conferences. After this IAUS 340, I am not likely to attend many more solar physics conferences. I shall certainly miss meeting you people at conferences. As the time comes for the curtain to fall, it is difficult to part ways with solar physics – my mistress of four decades. The Bible says: "Dust thou art, and unto dust shalt thou return." Thanks to modern astrophysics, we now know that the dust we are made of came from an ancient supernova. Thus, our life cycle is linked to the life cycle of the universe. I am privileged that I have seen some giants of science like Chandrasekhar and Parker fairly closely. They will inspire generations to come. For the rest of us average scientists, we come from darkness and again fade into darkness. Between these darknesses, we have a sunny interlude when we work and play. That is our life, that is our destiny.  I am happy that this sunny interlude has been so wonderful for me and I have received so much love and affection.  Thank you!

# Messages from well-wishers

*From Eugene Parker, University of Chicago:*

Let me take this festive occasion to congratulate you on a long and distinguished research career. I remember your early discussion with Chandra on the cultural differences between scientists in the east and the west. Chandra was initially vexed with your analysis but soon admitted that you had a valid point. Only a great scientist like Chandra would recognize the validity of the "upstart" view of an "upstart" student. Those were great days and you did not waste any time in getting on with your research once you had your degree.

Your treatise THE PHYSICS OF FLUIDS AND PLASMAS has proved to be a classic, of which you can be proud. And of which I can be proud that you were once my student who then moved on to another book and a long distinguished research career. We all salute you.

*From Eric Priest, University of St. Andrews:*

Ever since I first met you as a student in Chicago in the early 80s, I have watched your development into a world-class scientist and have admired greatly your fantastic advances in dynamo theory and your wonderfully fluent books. You have helped me understand many intricacies of the subject during our many discussions, but have also taught me the importance of a gracious and positive attitude to others. I wish you every happiness and many new ideas in future.

*From Paul Charbonneau, University of Montreal:*

Being absent in body but certainly not in thought, I wish to convey to you my warmest greetings and congratulations at this much deserved celebration of your life and work. . . so far!

Through your work, through the students you trained, through your teaching, your papers, your books, your impact in our field has been immense and I am sure will continue to be felt for a great many solar cycles.

From halfway across the world, we all here salute you!

*From Boon Chye Low, High Altitude Observatory:*

I join your friends and colleagues in India and from around the world in celebrating your professional scholarship and scientific accomplishments and contributions in our study of the Sun and astrophysical hydromagnetics.

We were both privileged to have received our education from Gene Parker and started our research with him. We learned much from Gene by just watching him in action and through the many things we talked about. You and I were fortunate with the professional opportunities that allowed us to determine our respective intellectual homes where we grew scientifically with our co-workers and by the works we did over the decades.

I remember you perceived that I was discouraging in your decision to return to India. I believe your decision was right and I was wrong in my discouragement. It seems to me that a decision sets the direction but does not guarantee the outcome. In your case, the outcome is extremely positive for you as scientist and for India at a time when advancement in society and economy needed solid leadership to educate her generations of new scientists and to lead in India's own scientific efforts.

I have much admiration for you. You must have inspired many at home and abroad with doing and accomplishing science as you would have in the US. It seems HAO needed you more than you needed HAO, a thought that came to me now and then over the years.

When I came to Chicago for graduate school in 1968, I did not plan to live in the US. . . The years I was at HAO were long enough for me to see major changes and consequences. I have much thoughts about scientific research and culture and about institutions and human foibles, from the perspective of my background.

Keep good health, Arnab. One day I will swing by on my way to see you and Ramit and have a few good conversations.